\documentclass[12pt,preprint]{aastex}
\shorttitle{Two Components and Their Decoupling in an EUV wave}
\shortauthors{Dai et al.}

\begin{document}

\title{Quadrature Observations of Wave and Non-Wave Components and Their Decoupling in 
an Extreme-Ultraviolet Wave Event}

\author{Y.~Dai\altaffilmark{1,2}, M.~D.~Ding\altaffilmark{1,2}, P.~F.~Chen\altaffilmark{1,2}, 
and J.~Zhang\altaffilmark{3}}

\altaffiltext{1}{School of Astronomy and Space Science, Nanjing University, Nanjing 210093, China}

\altaffiltext{2}{Key Laboratory of Modern Astronomy and Astrophysics (Nanjing University), 
Ministry of Education, Nanjing 210093, China}

\altaffiltext{3}{School of Physics, Astronomy, and Computational Sciences, George Mason University,
Fairfax, VA 22030, USA}

\email{ydai@nju.edu.cn}

\begin{abstract}
We report quadrature observations of an extreme-ultraviolet (EUV) wave event on 2011 January 27 obtained by 
the Extreme Ultraviolet Imager (EUVI) onboard \emph{Solar Terrestrial Relations Observatory}
(\emph{STEREO}), and the Atmospheric Imaging Assembly (AIA) onboard the \emph{Solar Dynamics Observatory}
(\emph{SDO}). Two components are revealed in the EUV wave event. A primary front is launched with an initial 
speed of $\sim$440 km s$^{-1}$. It appears significant emission enhancement in the hotter channel
but deep emission reduction in the cooler channel. When the primary front encounters a large coronal
loop system and slows down, a secondary much fainter front emanates from the primary front with a 
relatively higher starting speed of $\sim$550 km s$^{-1}$. Afterwards the two fronts propagate independently 
with increasing separation. The primary front finally stops at a magnetic separatrix, while the secondary
front travels farther before it fades out. In addition, upon the arrival of the secondary front, transverse
oscillations of a prominence are triggered. We suggest that the two components are of different natures. 
The primary front belongs to a non-wave coronal mass ejection (CME) component, which can be reasonably
explained with the field-line stretching model. The multi-temperature behavior may be caused by 
considerable heating due to the nonlinear adiabatic compression on the CME frontal loop. For the secondary
front, most probably it is a linear fast-mode magnetohydrodynamic (MHD) wave that propagates 
through a medium of the typical coronal temperature. X-ray and radio data provide us with complementary evidence
in support of the above scenario.  
\end{abstract}


\section{INTRODUCTION}
One of the most intriguing phenomena discovered by the Extreme-ultraviolet (EUV) Imaging Telescope 
\citep[EIT;][]{Delab95} onboard the \emph{Solar and Heliospheric Observatory} (\emph{SOHO}) 
satellite is ``EIT waves", which are characterized by a diffuse bright front globally
propagating through the solar corona \citep{Moses97,Thompson98}.  EIT waves were initially interpreted
as a fast-mode magnetohydrodynamic (MHD) wave in the corona \citep{Thompson99}, which can travel 
across the magnetic field lines freely, covering a quite large fraction of the solar disk. If the coronal 
fast-mode wave is strong enough, it can also perturb the much denser chromosphere at its base to produce 
an H$\alpha$ Moreton wave, just as the scenario proposed by \citet{Uchida68}. Many subsequent numerical and 
observational studies \citep[e.g.,][]{Wang00,Wu01,Warmuth04,Veronig06,Long08,Gopalswamy09,Patsourakos09a} 
have provided further evidence for this view. 

Such a fast-mode wave model was first challenged by \citet{Delannee99} who found that an EIT wave stopped 
at the magnetic separatrix, which is hard to explain in the wave framework. In addition, case studies have 
revealed that the EIT wave front is co-spatial with the coronal mass ejection (CME) frontal loop 
\citep[e.g.,][]{Attrill09,Chen09,Dai10}. Hence several alternative models have been proposed, which 
regard EIT waves as a result of magnetic reconfiguration related to the CME liftoff rather than a true 
wave in the corona. These non-wave models include the current shell model \citep{Delannee00}, the 
field-line stretching model \citep{Chen02,Chen05}, and the successive reconnection model \citep{Attrill07}. 
Besides, some other authors claim EIT waves to be a type of slow-mode MHD wave \citep{Wills-Davey07,Wang09}. 
For more details on the observations and modeling of EIT waves, please refer to recent reviews 
\citep{Wills-Davey09,Gallagher11,Zhukov11,Chen11b,Patsourakos12}.

\citet{Chen02} predicted that there should be a fast-mode wave ahead of the EIT wave, which was confirmed
by \citet{Harra03}. On the other hand, \citet{Zhukov04} suggested from the observational point of view 
that there could be both wave and non-wave components in an 
EIT wave. However, early EIT wave studies to catch such multiple components often suffered the low 
cadence of EIT, which is 12 minutes at best. The situation has been greatly improved with the launch 
of the \emph{Solar Terrestrial Relations Observatory} \citep[\emph{STEREO};][]{Kaiser08} and the
\emph{Solar Dynamics Observatory} \citep[\emph{SDO};][]{Pesnell12}. Thanks to the much higher temporal 
resolutions of the EUV telescopes onboard the three spacecraft, multiple components in an EIT wave have been
successfully identified in observations \citep[e.g.,][]{Liu10,Chen11a,Cheng12,Asai12} and verified in 
numerical efforts \citep[e.g.,][]{Cohen09,Downs11,Downs12}. With the observations of modern generation of 
EUV imagers, now we prefer the more general term ``EUV wave" to the conventional one ``EIT wave". 
In this paper we report quadrature observations of two components and their decoupling in an EUV wave event on
2011 January 27 from both \emph{STEREO} and \emph{SDO}\@. The distinct differences in amplitude, kinematics,
and multi-temperature behavior imply their different physical mechanisms. In Section 2 we introduce the instruments 
and data sets. Analysis is carried out and results are presented in Section 3. Then we discuss the results
in Section 4 and draw our conclusions in Section 5.

\section{INSTRUMENTS AND DATA SETS}
The EUV wave under study was launched on 2011 January 27 around 12:00 UT from NOAA active region (AR)
11149 when the AR was very close to the northwest limb from the Earth perspective. At that time the 
\emph{STEREO Ahead} satellite (\emph{STEREO-A}) was $\sim86\degr$ west of the Earth. Therefore the 
location of the source region and the quadrature configuration of \emph{STEREO-A} and near-Earth
\emph{SDO} offer us a perfect opportunity to trace the evolution of the EUV wave both face-on 
(from \emph{STEREO-A}) and edge-on (from \emph{SDO}). 

We used EUV imaging data from the Extreme Ultraviolet Imager \citep[EUVI;][]{Wuelser04} 
onboard \emph{STEREO} and the Atmospheric Imaging Assembly \citep[AIA;][]{Lemen12} onboard \emph{SDO}\@.
EUVI, part of the Sun Earth Connection Coronal and Heliospheric Investigation \citep[SECCHI;][]{Howard08} 
instrument suite, observes the chromosphere and corona up to 1.7 $R_{\sun}$ in 4 EUV channels with a pixel 
size of 1$\farcs$58. AIA provides multiple simultaneous images of the transition region and corona up to 
1.5 $R_{\sun}$ in 10 EUV and UV channels with 0$\farcs$6 pixel size and 12-second temporal resolution. 
In this work, we focused on the \emph{STEREO-A}/EUVI (hereafter EUVI-A) 195~\AA\ and AIA
171, 193, and 211~\AA\ observations for the reason that in general, EUV waves are best observed 
at these wavelengths \citep[cf.,][]{Veronig08,Li12}. During the period of interest the cadence of 
the EUVI-A 195~\AA\ channel was 5 minutes.

\section{ANALYSIS AND RESULTS}

\subsection{EVOLUTION OF THE EUV WAVE}
We used base ratio images to study the wave evolution. Images were first prepared and differentially 
rotated to the same pre-event time at 11:50 UT using the standard IDL routines in Solar Software. 
Then an image taken around 11:50 UT was selected as the reference image for each channel; all the following
images were  divided by the  corresponding reference images. Figure 1 and the associated online 
Animation 1 show the on-disk evolution of the EUV wave in EUVI-A 195~\AA\@. The eruption site is located on the
southern side of AR11149. Due to the great magnetic gradient to the north, the EUV wave propagates mainly 
southward instead of isotropically. First observed at 12:00 UT, the wave front initially expands very fast.
By 12:05 UT, it has been fully developed, appearing as a diffuse bright rim that covers an angular span 
over 110$\degr$ (Figure 1$b$). Dimming regions are seen following the expanding wave front. Afterwards, 
the bright wave front undergoes a significant deceleration, especially in the south direction, and finally
stops on the southern hemisphere, forming a stationary bright stripe along the latitudinal direction 
(Figures 1$e$--$f$). As the bright wave front slows down, another much fainter front emanates and propagates
ahead of it, attaining a distance far beyond the stationary front (Figures 1$c$--$f$ and Animation 1). 
However, due to the relatively low cadence and sensitivity of EUVI as well as the nature of the base ratio 
method, this wave signal weakens so quickly that its evolution cannot be reliably traced in EUVI-A\@. 

In order to investigate the wave kinematics in EUVI-A in an objective manner, we adopted the 
semi-automated detection algorithm described in \citet{Long11} to identify and track the bright wave front. 
We selected a wave sector extending from the eruption center (-100$\arcsec$, 300$\arcsec$) in direction 
$180\degr\pm5\degr$ (directly southward), within which the wave kinematics was studied. A perturbation 
profile was derived by averaging the base ratio intensity values in annuli of 1$\degr$ width with 
increasing radii on the spherical solar surface. At each observation time, the perturbation profile was 
fitted with a Gaussian curve, of which the peak position was taken as the distance of the wave front. 
Figure 2$a$ illustrates such a Gaussian fit to the perturbation profile of the EUV wave at 12:05 UT\@. 
It is worth noting that the intensity enhancement of the EUV wave at that time is as high as 80\%. 
The distinct deceleration of the bright wave front is validated by the wave kinematics shown in Figure 2$b$.
The wave front decelerates from an initial speed of 398 km s$^{-1}$ to zero velocity within a 
period of 20 minutes. Eventually it turns into a stationary front at a distance $\sim$500 Mm south to the 
eruption center.

Online Animations 2--4 show the limb evolution of the EUV wave in AIA 211, 193, and 171~\AA,
respectively. Some snapshots of the animations are picked to display in Figure 3. A front appears
around 12:00 UT (Figure 3$a$), and strengthens quickly into a diffuse bright front in 211~\AA\ and 
193~\AA\ (Figures 3$b$ and $g$). However, in 171~\AA, the main body of the front appears dark (Figure 3$i$).
The front is largely inclined to the limb, so in the early stage it propagates mainly laterally
rather than radially. Thanks to the extremely high cadence and sensitivity of AIA as well as a lower
background with less contribution from the disk, the emanation and separation of a secondary faint front 
from the primary front are revealed when the primary front encounters a large coronal loop system 
(clearly seen in 171~\AA) and then slows down (Animations 2--4). Afterwards, the two fronts evolve 
independently. The primary front decelerates significantly and finally stops (Figures 3$f$, $h$, and $i$), 
while the secondary front travels farther as it gradually fades out (Animations 2--4). 

To avoid any ambiguities introduced from close-to-limb disk regions, we studied the off-limb wave behavior
at a heliocentric height of 1.1 $R_{\sun}$ (the black circle in Figure 3), since there is mounting 
evidence that EUV waves are confined to a region 1--2 scale heights above the chromosphere 
\citep[e.g.,][]{Patsourakos09b}. Along the circle we actually traced the evolution of the EUV wave in nearly 
the same direction as that selected in EUVI-A\@. Figure 4 shows the time-position angle (PA) diagrams of 
the EUV wave in AIA 211, 193, and 171~\AA, respectively. It is clearly seen that the kinematics of the EUV wave is
almost the same among different channels, with the red and blue lines visually tracking the primary and secondary 
fronts, respectively. 

We converted the PA values to distances to the eruption center (at a PA of $\sim282\degr$) and then redrew the trajectories of the primary and secondary fronts in Figure 5$a$. For comparison, we over-plotted the 
time-distance data of the bright wave front in EUVI-A 195~\AA, which were multiplied by a factor of 1.1 to 
compensate for the difference in tracing heights (1.1 $R_{\sun}$ for AIA versus 1.0 $R_{\sun}$ for EUVI-A).
As expected, the kinematics of the primary front in AIA is in perfect agreement with that of the bright wave front 
in EUVI-A, indicating that these two fronts are the same feature but viewed from different perspectives.  
The velocity evolution of the primary and secondary fronts is displayed in Figure 5$b$. The primary front exhibits 
an initial speed of 443 km s$^{-1}$ and undergoes only a slight deceleration in the early stage. At 12:07 UT, the exact
time when the primary front interacts with the large coronal loop system south of it and starts to decelerate 
significantly, the secondary front emanates from the primary front with a higher starting speed of 
553 km~s$^{-1}$. Since then the separation of the two fronts has been increasing, leading to the decoupling of the two 
fronts. The velocity of the primary front finally decreases to zero, and the secondary
front also decelerates considerably before its strength quickly drops below the detectable level. We should bear
in mind that the kinematic analysis for the secondary front is subject to much more uncertainties than that for 
the primary front due to its much fainter appearance. Although lacking quantitative comparisons, we believe that
the secondary front in AIA corresponds to the very weak wave signature in EUVI-A\@. 

As can be also seen in Figure 4, the two fronts in AIA show different
emission patterns. For the primary front, it exhibits prominent emission enhancement in 211~\AA, moderate 
enhancement in 193~\AA, but deep depletion in 171~\AA\@. Emission reduction of the wave front in 
171~\AA\ was previously reported \citep[e.g.,][]{Dai10,Liu10}. For the secondary front, it is the 
strongest in 193~\AA, relatively weaker in 211~\AA, and nearly invisible in 171~\AA\@.

\subsection{ASSOCIATED PHENOMENA}
Associated with the EUV wave, there is a \emph{GOES} C1.2 class flare. The \emph{GOES} 1--8 \AA\ soft X-ray (SXR)
light curve in Figure 6$a$ indicates that the flare takes place between 11:53 UT and 12:05 UT, with the peak time at
12:01 UT\@. During the event time, the \emph{RHESSI} satellite was affected by the South Atlantic Anomaly (SAA). Thus 
we used the derivation of the \emph{GOES} SXR light curve shown in Figure 6$b$ as a proxy of the hard X-ray (HXR)
evolution of the flare. The so derived HXR light curve also peaks at around 12:01 UT, slightly earlier than the 
SXR peak. Both the SXR and HXR light curves indicate that this is an impulsive flare. By the peak time of the flare, 
the primary front has been formed at a large distance, implying that the impulsive flare pulse occurs too late to 
drive the EUV wave event.

Radio observations from the Radio Solar Telescope Network (RSTN; 25--180 MHz) in the period of interest are displayed 
in Figure 6$c$ as dynamic spectrum in the metric domain. Besides a type III burst that coincides with a small 
HXR spike at 11:59 UT, the dominant feature is a type II burst starting from 12:08 UT, with a staring frequency of 
83 MHz at the harmonic band. The occurrence of the metric type II burst follows the decoupling of the primary and
secondary fronts within one minute, which may reflect a physical link between the decoupling process and a coronal
shock. However, when assuming a coronal density model for the quite Sun at solar minimum, which was proposed by 
\citet{Saito77}, the coronal shock inferred from the type II burst starts at a heliocentric height over
1.4 $R_{\sun}$, significantly higher than the detectable altitude of the secondary front. In addition, the signal 
of the secondary front is very weak. Therefore, if the secondary front is a part of the coronal shock, it must
be away from the nose of the shock.

The EUV wave also triggers transverse oscillations of a prominence over the southwestern limb. Figure 7$a$ displays
the prominence morphology in AIA 193 \AA, which appears as a dark feature at a PA of 248$\degr$. We studied the 
prominence oscillations along the azimuthal direction (the white slice in Figure 7$a$). As shown in Figure 7$b$, the 
transverse oscillations start from 12:10 UT, with the multiple prominence threads first moving southward and then 
moving northward. The oscillation period is about 14 minutes, and the maximum amplitude is about 8000 km. Compared
with the wave kinematics, the start of the prominence oscillations coincides with the arrival of the secondary
front, which can be further validated by the bright features at 12:10 UT in Figure 7$b$. This observational 
factor may indicate a wave nature of the secondary front. Recently, \citet{Asai12} and
\citet{Liu12} also observed prominence transverse oscillations triggered by limb EUV waves. The oscillation 
parameters in our study are consistent with those in \citet{Asai12}. In \citet{Liu12}, the prominence oscillations 
last for a longer interval, with oscillation periods about twice longer. Nevertheless, the physics that determines 
the oscillation parameters is beyond the scope of this paper.

\section{DISCUSSION}
We report the \emph{STEREO-A}/EUVI and \emph{SDO}/AIA quadrature observations of the EUV wave event on 2011 
January 27, in which two fronts and their decoupling are revealed. From the edge-on perspective of AIA, 
the wave fronts extend to a quite high altitude, implying that the kinematics analysis from the
single face-on perspective of EUVI-A would somewhat underestimate the wave speed owing to the lack of 
knowledge on the height of the line-of-sight integration maximum \citep{Kienreich09}. Therefore, the value
of $\sim$440 km s$^{-1}$ measured 0.1 $R_{\sun}$ above the limb may reflect a more real initial speed of 
the primary front. The first appearance of the primary front occurs earlier than the peak of the associated
impulsive flare, which invalidates a flare driver of the EUV wave event.

The primary and secondary fronts show distinct differences in amplitude, kinematics, and multi-temperature behavior,
which imply their different physical mechanisms. In Figure 8 we show the coronal magnetic topology close to the
event time, which was extrapolated from the \emph{SOHO}/Michelson Doppler Imager \citep[MDI;][]{Scherrer95}
synoptic magnetogram with the potential-field source-surface \citep[PFSS;][]{Schrijver03} model. The
extrapolated magnetic field lines are overlaid on the simultaneous base ratio images of the EUV wave in 
AIA 193~\AA\ and EUVI-A 195~\AA\ at 12:25 UT when the primary front has turned into a stationary front. 
The magnetic topology shows a large-scale magnetic system that covers an extent from AR 11149 to the
elongated magnetic separatrix on the southern hemisphere. It is clearly seen that the stationary front is 
indeed co-spatial with the magnetic separatrix, indicative of a non-wave nature of the primary front. 
On the contrary, the secondary front triggers transverse oscillations of a prominence and travels across the 
magnetic separatix to a further distance, which are typical characteristics of fast-mode waves. 

Seen from the AIA limb observations, the primary front extends continuously down to the limb, which might
not be explained by the current shell model \citep{Delannee00} in which the brightening due to Joule heating
is confined quite high in the corona. Lack of detailed small-scale magnetic topology makes us unable to judge 
if the successive reconnection model \citep{Attrill07} works for this event. Instead, the field-line stretching mode 
\citep{Chen02,Chen05} seems to be a reasonable explanation. In this model, the primary front corresponds 
to the CME frontal loop that is composed of the newly stretched magnetic field lines. Guided by the 
overlying large-scale magnetic system, in the early stage the CME frontal loop propagates with a substantial
inclination toward the limb, showing a fast lateral expansion. Meanwhile, a large amount of material is
quickly piled onto the frontal loop, resulting in a nonlinear density enhancement (Figure 2$a$, assuming a
wide temperature coverage of the EUVI 195~\AA\ channel). Furthermore, this adiabatic compression process
leads to considerable heating. The heating effect makes further positive contribution to the emission 
enhancement in the hotter AIA 211~\AA\ channel (with $T_{\mathrm{peak}}$ of $\sim$2 MK) in addition to the 
density enhancement (Figure 4$a$). In the AIA 193~\AA\ channel ($T_{\mathrm{peak}}$ $\sim$1.3 MK, a typical
coronal temperature), such contribution may not be so significant, or it could be even somewhat negative 
(Figure 4$b$). For the cooler AIA 171~\AA\ channel ($T_{\mathrm{peak}}$ of $\sim$0.6 MK), the response 
function decreases very fast from the peak with increasing temperatures. Therefore, in 171~\AA, the heating
strongly reduces the emission (Figure 4$c$), and the density enhancement cannot compensate for the emission decrease 
caused by the temperature rise. According to the field-line stretching model, the CME can only stretch the magnetic
field lines of the same magnetic system within which the CME is involved. At the magnetic separatrix, 
a border with other magnetic systems, the CME frontal loop stops and forms the stationary front. 
It is worth noting that an associated CME is later observed in the high 
corona (see \url{http://spaceweather.gmu.edu/seeds/lasco.php}), whose southern border is roughly 
located at the PA of the magnetic separatrix.
 
It is believed that the CME has driven a fast-mode wave since it starts the lateral expansion. 
However, this fast-mode wave is not distinguishable from the CME until the CME frontal loop encounters the 
large coronal loop system south of it. The interaction between the CME and the coronal loop system not 
only slows down the CME lateral expansion, but also increases the local fast-mode speed. As a result, 
the fast-mode  wave (the secondary front) emanates from the CME frontal loop with a relatively higher
``starting" speed ($\sim$550 km s$^{-1}$). From then on, the fast-mode wave is decoupled from the CME 
and the two components evolve independently. The CME changes its propagation from mainly in the lateral
direction to mainly in the radial direction. As the CME propagates radially outward, the Alfv\'{e}n speed
first increases to a maximum and then decreases, facilitating the formation of a CME-driven shock at a relatively 
high altitude. This could be a reasonable explanation to the metric type II burst in this 
work. We note that the case study of a 
coronal shock by \citet{Gopalswamy12} shows that the Alfv\'{e}n speed attains a maximum of $\sim450$ km s$^{-1}$ at
a heliocentric height of $\sim$1.35 $R_{\sun}$. For the fast-mode wave, it travels across the magnetic field lines
freely, and triggers the prominence transverse oscillations over the southwestern limb. As the fast-mode wave 
propagates into quite Sun regions, the  decrease in magnetic strength leads to the wave deceleration. Compared 
to the CME frontal loop, the fast-mode wave is much fainter. In addition, the wave signature is stronger in 
193~\AA\ than that in 211~\AA, and almost invisible in 171~\AA\@. Combining these observational facts 
together, we suggest that the fast-mode wave is a linear MHD wave that propagates through a medium of the 
typical coronal temperature. 

As mentioned above, there have been several observational studies dealing with EUV wave events with
two fronts and their decoupling. In \citet{Cheng12}, they found that the lateral expansion of the CME 
bubble first accelerates and the diffuse front is separated from the CME bubble  shortly after the lateral 
expansion slows down. In their case, the associated flare is rather gradual, and the acceleration of the CME 
coincides with the flare's rising phase. In our study, the associated flare is an impulsive one, so the CME may 
undergo a very impulsive acceleration in its initiation phase \citep[cf.,][]{Zhang01}. As a result, upon its first
appearance, the CME lateral expansion (primary front) has already attained a maximum speed of $\sim$440 km~s$^{-1}$. 
Furthermore, the lateral expansion of the CME bubble in \citet{Cheng12} should reflect an intrinsic
expansion of the CME, while in our case the CME lateral expansion is mainly guided by the overlying large-scale
magnetic system. The event studied by \citet{Asai12} is a very intense one, in which an H$\alpha$ Moreton wave is
observed co-spatial with the sharp bright EUV wave front in the very early stage. This implies that at the
very beginning, the major CME has driven a coronal MHD wave initially strong enough to penetrate to 
the chromosphere, which is further validated by a concurrent metric type II burst. As the bright EUV wave 
front (which we believe corresponds to the CME frontal loop) decelerates to an ``ordinary EIT wave", the MHD 
wave is detached from the CME and its strength decreases to the linear regime, unable to perturb the chromosphere
any more. However, in our study, the secondary front keeps a linear MHD wave during its whole evolution process. 
If the secondary front is a part of the coronal shock that starts shortly after the decoupling of the two fronts, 
it must be away from the nose of the shock where the wave strength is the strongest. In \citet{Chen11a}, they also 
observed that the slow wave front finally stops at a magnetic separatrix. The event they studied is associated 
with a microflare, and no CMEs are detected during the event time. While in our study, an associated CME is later
observed, with the location of its southern border consistent with the PA of the magnetic separatrix. This 
supplies further evidence for the non-wave nature of the primary front.

Finally, we notice that the EUV wave is the second one of three homologous EUV wave events studied 
in \citet{Kienreich12}. They found that the wave is later reflected at the border of the extended 
coronal hole at the southern polar region. Hence they concluded that the EUV wave is purely a 
fast-mode wave. We think the reflected wave should correspond to the secondary front in our study, 
which is indeed a fast-mode wave. In the early stage, it is actually attached on the non-wave CME 
component.

\section{CONCLUSIONS}
By using the \emph{STEREO-A}/EUVI and \emph{SDO}/AIA quadrature observations of an EUV wave event on 2011 
January 27, two fronts and their decoupling are revealed. The two fronts show distinct differences in amplitude,
kinematics, and multi-temperature behavior. Complemented with the X-ray and radio observations, we suggest that 
the two fronts are of different natures. The primary front belongs to a non-wave CME component, which 
can be reasonably explained with the field-line stretching model. For the secondary front, most probably it is
a linear fast-mode MHD wave that propagates through a medium of the typical coronal temperature. The decoupling
of the two fronts is caused by the interaction of the CME frontal loop and a large coronal loop system south of it.

\acknowledgements{
We are grateful to the anonymous referee whose constructive comments led to a significant improvement
of the manuscript.
This work is supported by NSFC under grants 11103009, 10933003, 11078004, and 11025314, and by 
NKBRSFC under grant 2011CB811402. Y.D. is also sponsored by SRFROCS, Ministry of Education.
We thank the \emph{STEREO}/SECCHI, \emph{SDO}/AIA, \emph{GOES}, and RSTN consortia for their open data policy. }

%\bibliographystyle{apj}
%\bibliography{ms}

\clearpage
\begin{figure}
\epsscale{0.8}
\plotone{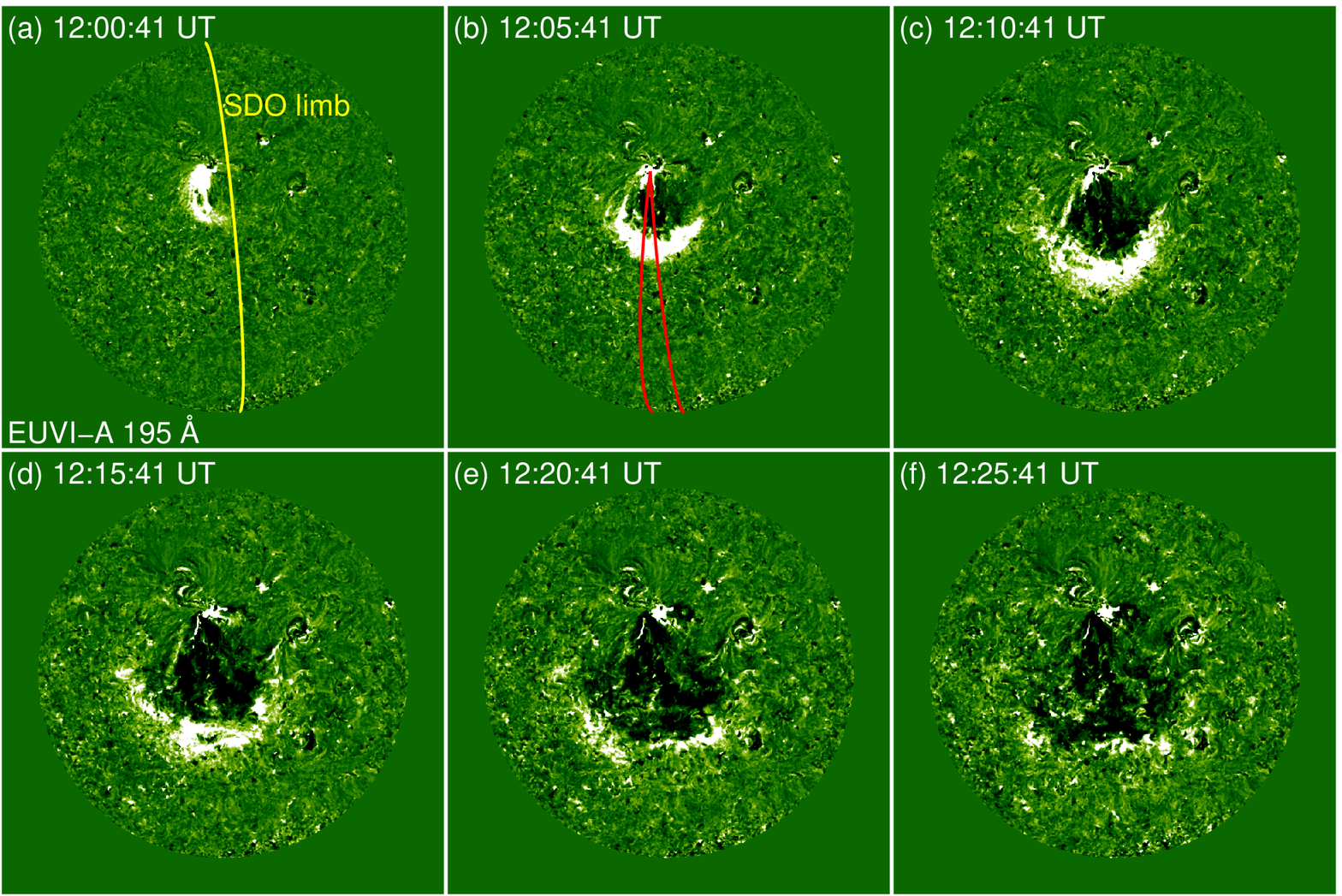}
\caption{Base ratio images of the EUV wave taken by \emph{STEREO-A}/EUVI at 195~\AA\@. 
The yellow line outlines the solar limb viewed from \emph{SDO}, and the red lines 
indicate great circles through the eruption center (-100$\arcsec$, 300$\arcsec$) that
border the wave sector in direction $180\degr\pm5\degr$ within which the wave kinematics is
studied. Note that all \emph{STEREO-A} observation times in this work are corrected to Earth UT to 
compensate for the slight difference in light travel times from the Sun to \emph{STEREO-A} and 
\emph{SDO}\@.}
\end{figure}

\clearpage
\begin{figure}
\epsscale{0.7}
\plotone{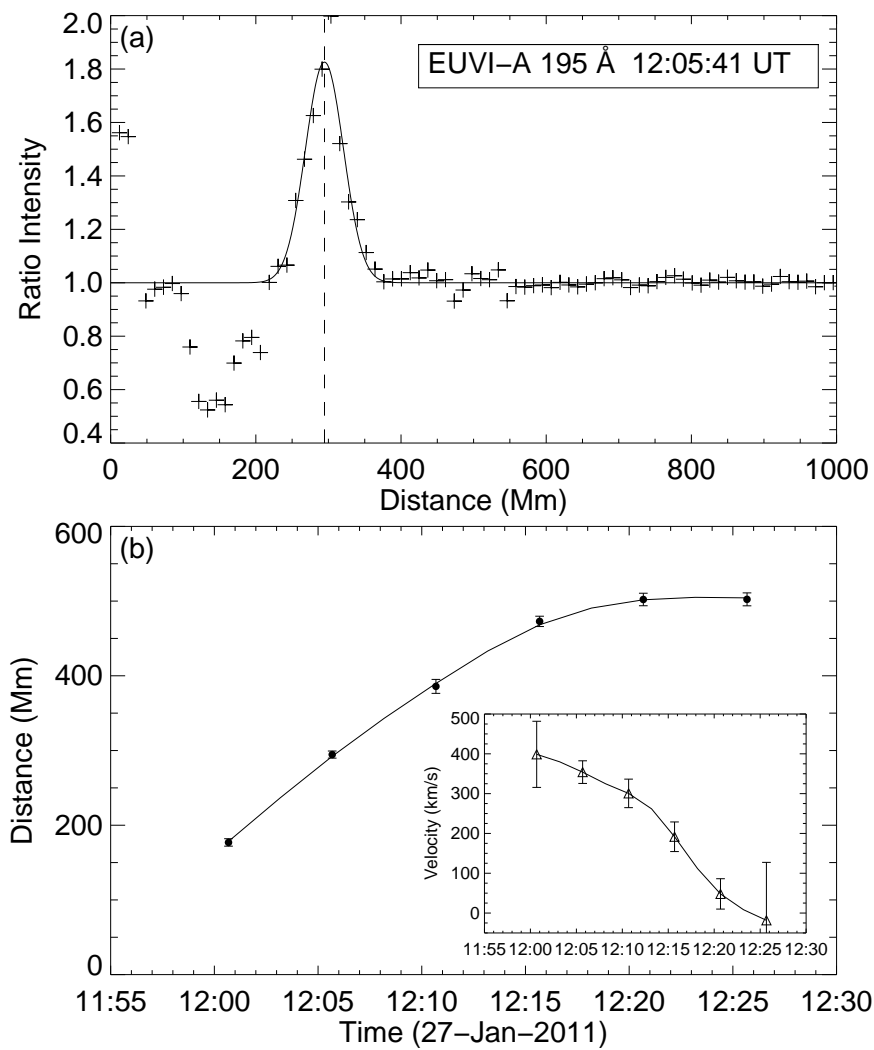}
\caption{\emph{Top}: Gaussian fit to the perturbation profile (plus signs) of the EUV wave at 
12:05:41 UT except for the flaring and deep dimming sections. The vertical dashed line marks 
the position of the Gaussian peak that represents the distance of the wave front at that time.
\emph{Bottom}: Time-distance diagram of the bright wave front (filled circles) together with the 
spline fit. Inset is the velocity evolution derived from differentiation to the fitted points (triangles) 
using 3-point Lagrangian interpolation.}
\end{figure}

\clearpage
\begin{figure}
\epsscale{0.9}
\plotone{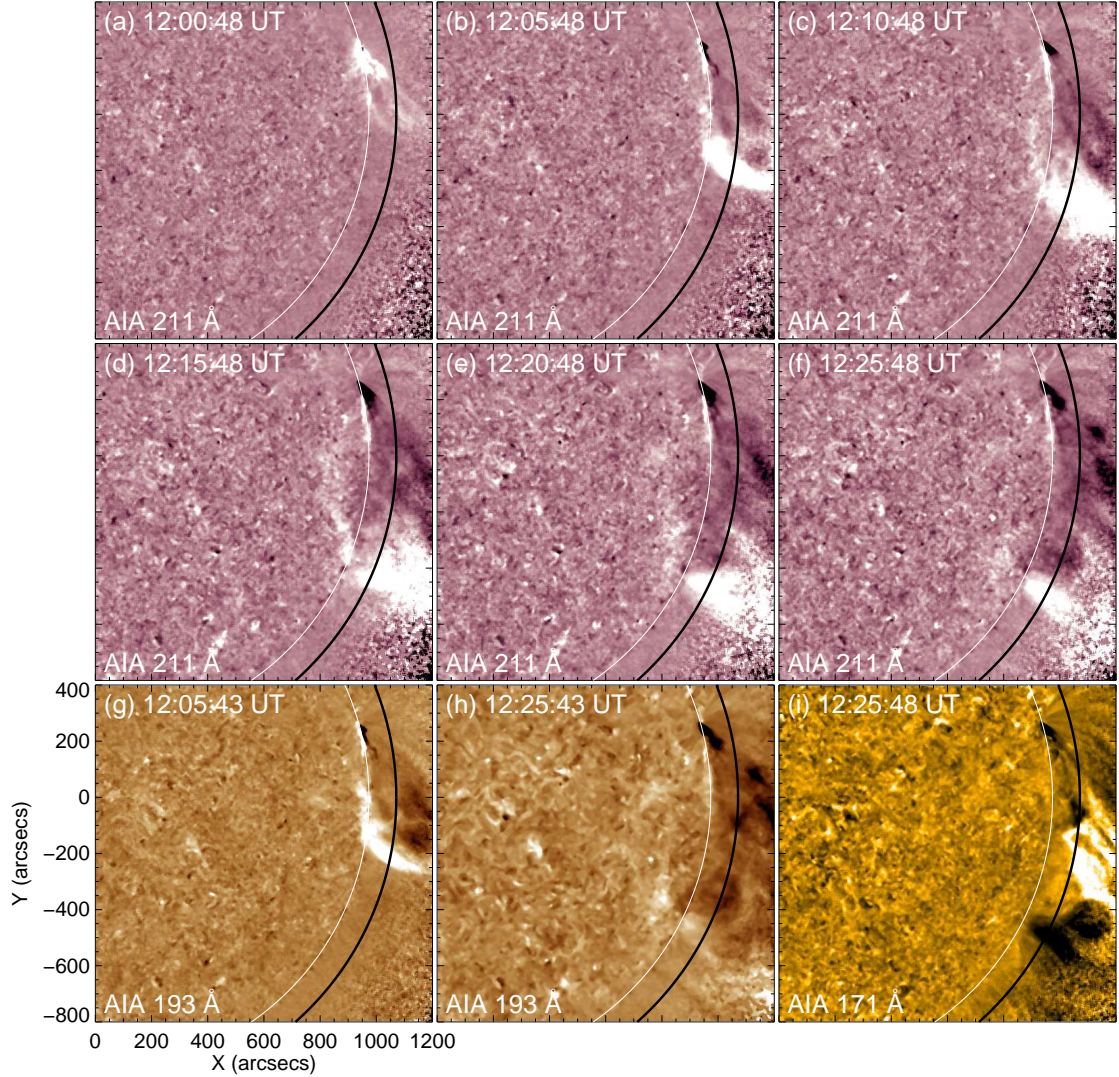}
\caption{Base ratio snapshots of the EUV wave taken by \emph{SDO}/AIA at 211 ($a$--$f$), 
193 ($g$--$h$), and 171~\AA\ ($i$), respectively. The black circle is located 0.1 $R_{\sun}$ 
above the limb along which the off-limb evolution of the EUV wave is traced.}
\end{figure}

\clearpage
\begin{figure}
\epsscale{0.6}
\plotone{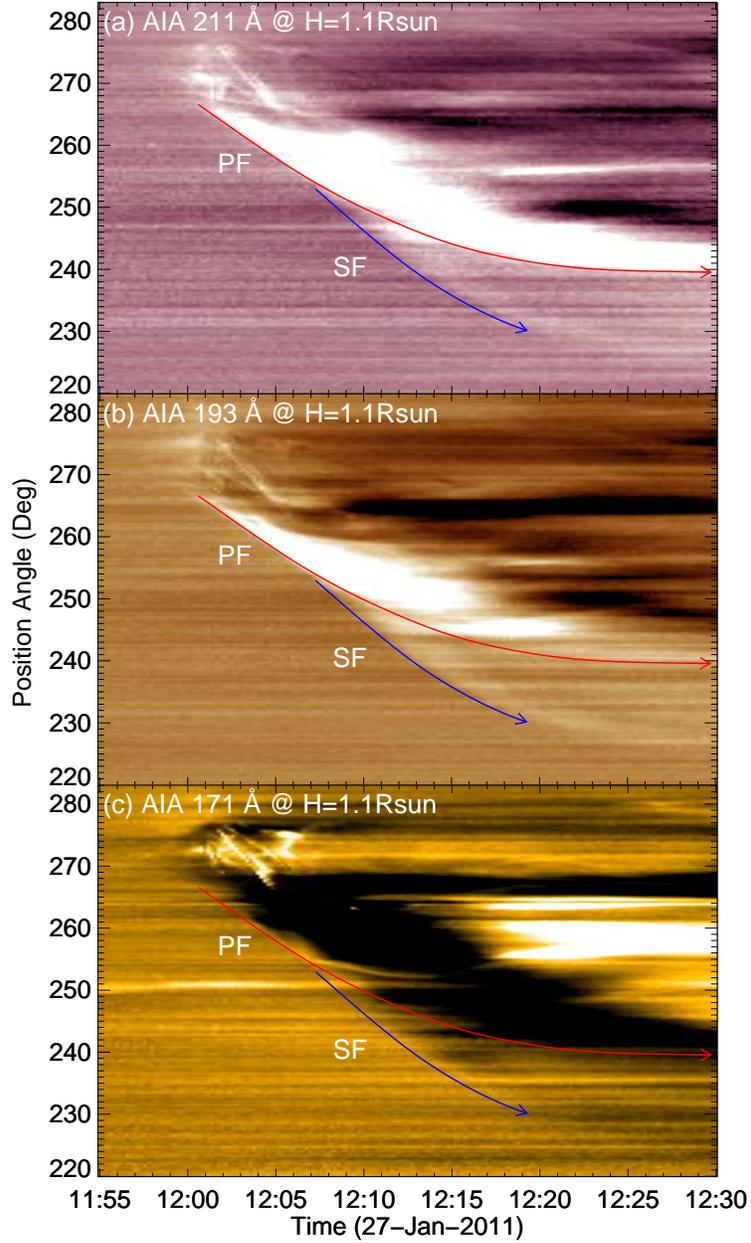}
\caption{Time-PA diagrams of the EUV wave in \emph{SDO}/AIA 211 ($a$), 193 ($b$), and 171~\AA\ ($c$), 
respectively. The red line (PF) follows the primary front, while the blue line (SF) tracks the secondary front. 
The PA values are counted counterclockwise from the solar north.}
\end{figure}

\clearpage
\begin{figure}
\epsscale{0.7}
\plotone{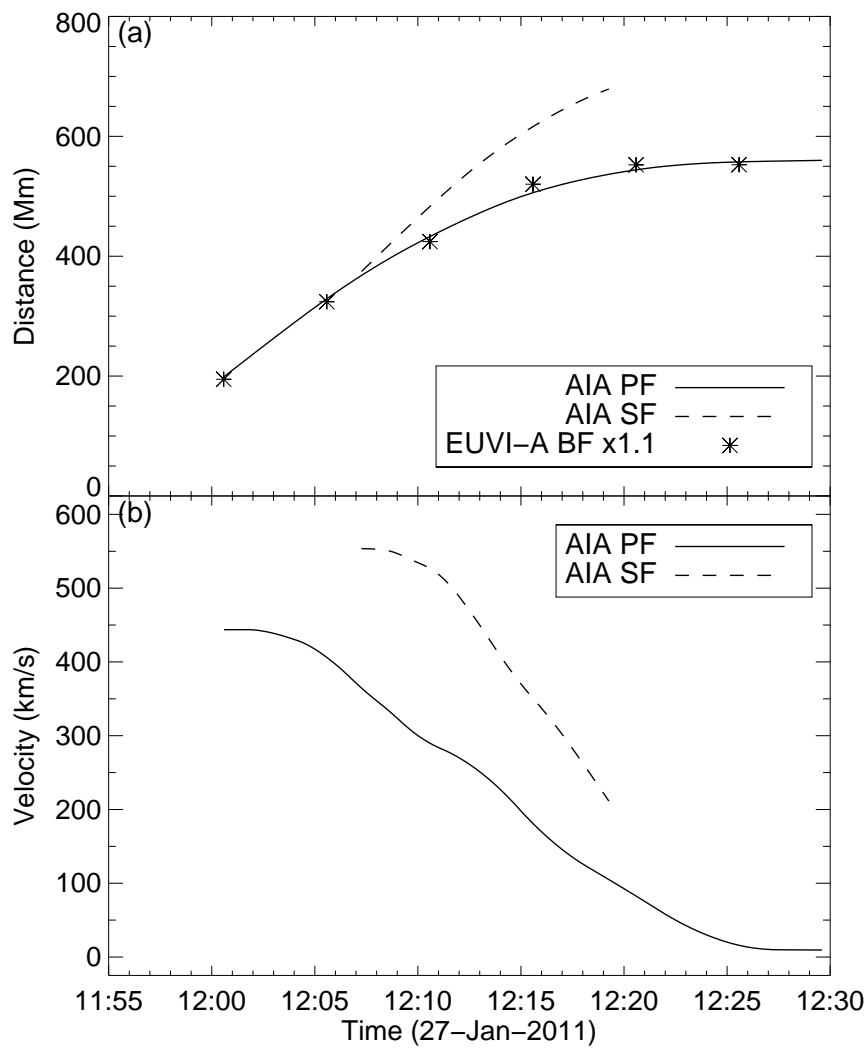}
\caption{\emph{Top}: Time-distance diagram of the primary (solid line) and secondary (dashed line) 
fronts in AIA\@. The over-plotted asterisks are the same time-distance data of the bright wave front in 
EUVI-A 195~\AA\ in Figure 2$b$ but multiplied by a factor of 1.1. \emph{Bottom}: The velocity evolution of the 
primary (solid line) and secondary (dashed line) fronts obtained using the same method as that for
the bright wave front in EUVI-A. Note that the cadence of the measurements for AIA is 12 seconds.}
\end{figure}

\clearpage
\begin{figure}
\epsscale{0.7}
\plotone{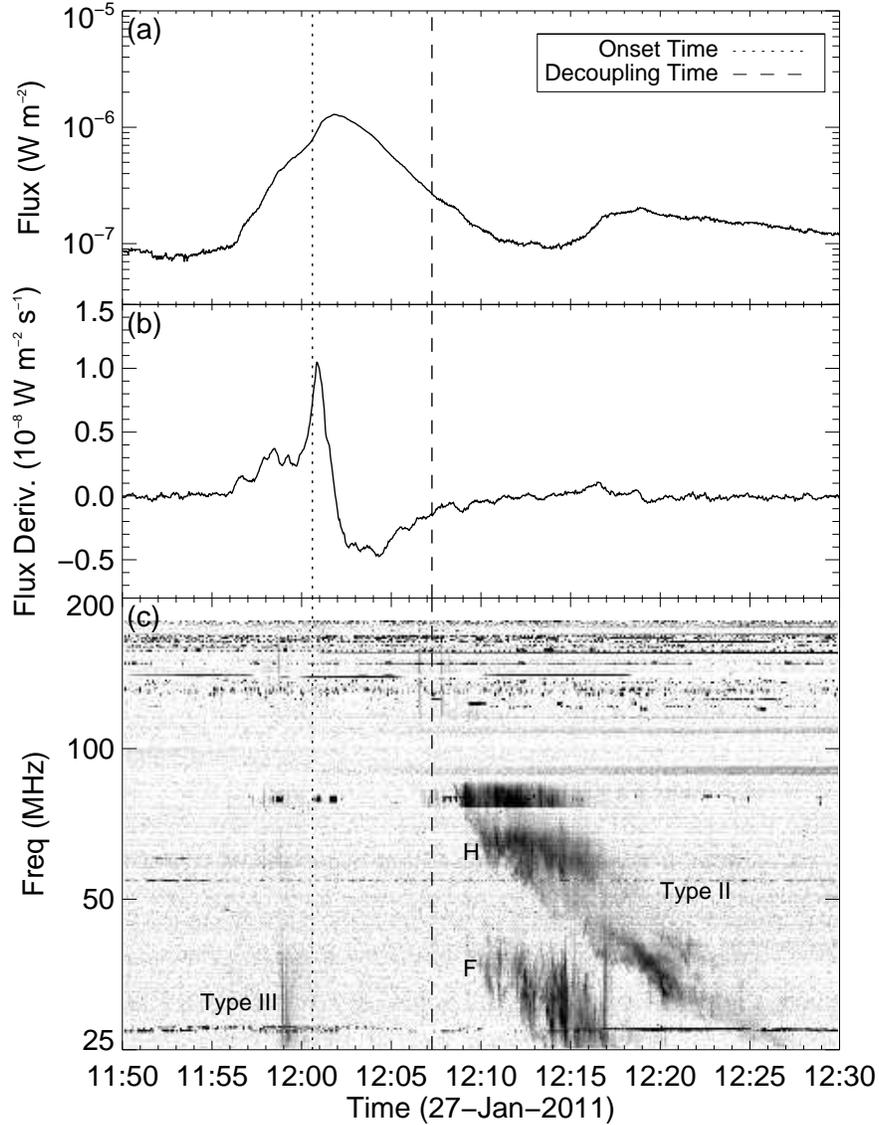}
\caption{X-ray and radio observations of the event. The top two panels show \emph{GOES} 1--8 \AA\ 
light curve ($a$) and its derivation smoothed by a boxcar of 30 s ($b$), respectively. The bottom panel 
shows radio dynamic spectrum obtained by RSTN (25--180 MHz), in which a type III burst and a type II burst are 
revealed. The letter ``F" (``H") denotes the fundamental (harmonic) band of the type II burst. The vertical dotted 
line indicates the onset time of the EUV wave, and the vertical dashed line represents the decoupling
time of the primary and secondary fronts.}
\end{figure}

\clearpage
\begin{figure}
\epsscale{0.9}
\plotone{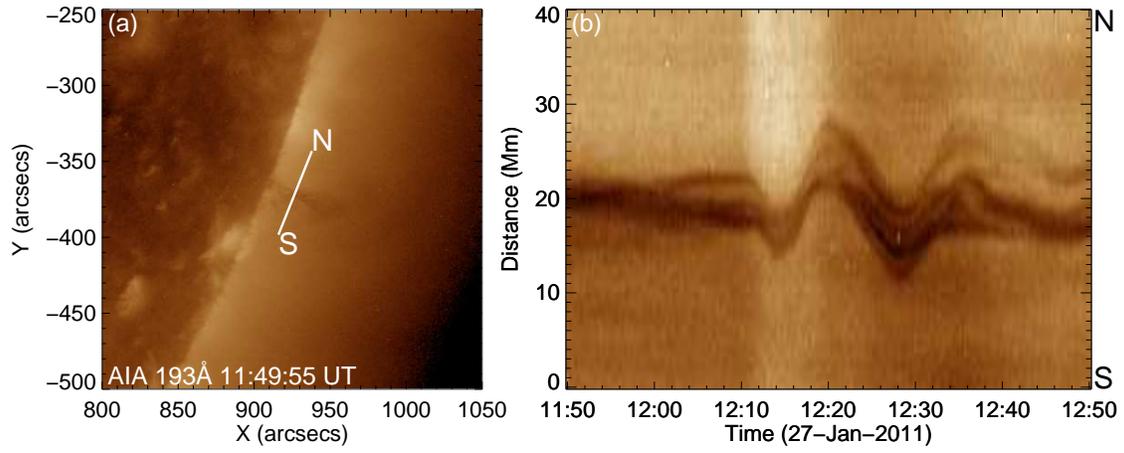}
\caption{\emph{Left}: Pre-event AIA 193 \AA\ image showing a prominence at a PA of 248$\degr$. 
\emph{Right}: Time-distance diagram of the oscillating prominence along the white slice outlined in the left panel.
The distances are measured from the south.}
\end{figure}

\clearpage
\begin{figure}
\epsscale{0.9}
\plotone{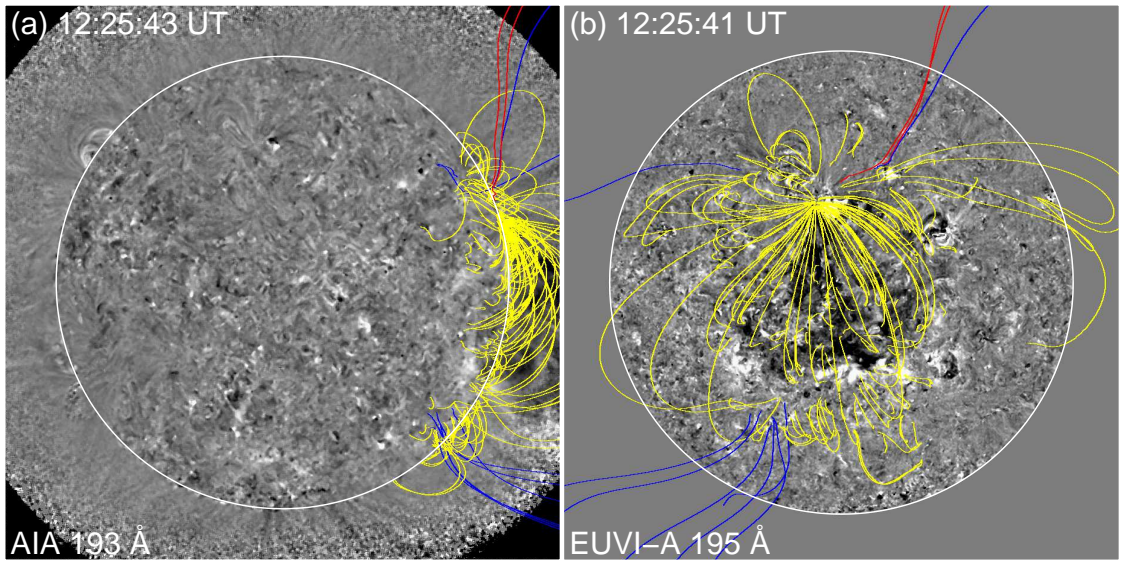}
\caption{Coronal magnetic topology close to the event time, which is extrapolated from the 
\emph{SOHO}/MDI data with the PFSS model, overlaid on the simultaneous base ratio images of the 
EUV wave in \emph{SDO}/AIA 193~\AA\ (\emph{left}) and \emph{STEREO-A}/EUVI-A 195~\AA\ (\emph{right}) 
according to each spacecraft's perspective. The yellow lines represent closed magnetic field 
lines, and the blue (red) lines denote open magnetic field lines from positive (negative) polarity.}
\end{figure}
\end{document}